\documentclass[aps,prd,column,showpacs,showkeys,amsmath,amssymb,nofootinbib]{revtex4}
\usepackage{mathrsfs}
\usepackage{graphicx}% Include figure filesph
\usepackage{dcolumn}% Align table columns on decimal point
\usepackage{bm}

\def\beq{\begin{equation}}
\def\eeq{\end{equation}}
\def\bea{\begin{eqnarray}}
\def\eea{\end{eqnarray}}

\def\fun#1#2{\lower3.6pt\vbox{\baselineskip0pt\lineskip.9pt
  \ialign{$\mathsurround=0pt#1\hfil##\hfil$\crcr#2\crcr\sim\crcr}}}
\begin{document}

\preprint{}

\title{Revisiting the assignment of $1^{3}D_{1}$ meson nonet}

\author{Xue-Chao Feng$^{1}$,Ke-Wei Wei$^{2}$ and Jie Wu$^{1}$ }
\affiliation{$^{1}$College of Physics and Electronic Engineering, Zhengzhou University of Light Industry, 450002 Zhengzhou, China}
\affiliation{$^{2}$School of Science,Henan University of Engineering, 451191 Zhengzhou, China}

\begin{abstract}
Based on the mass relations from the Regge phenomenology, we revisit the mass spectrum of $1^{3}D_{1}$ meson nonet. The masses of kaon and $s\bar{s}$ member of $1^{3}D_{1}$ ($1^{3}D_{1}(s\bar{s})$) meson nonet are obtained and the results are compared with the values from different theoretical models. Moreover, the strong decay properties of the kaon and $1^{3}D_{1}(s\bar{s})$ are presented in the $^{3}P_{0}$ model. On the basis of results, we suggest the assignment of $1^{3}D_{1}$ meson nonet need further tested in the new experiment in the future. Our results also provide mass constraints for the study of these states.
\end{abstract}


\maketitle
\section{Introduction}

Quantum Chromodynamics (QCD) is widely accepted as a non-Abelian gauge field theory to describe strong interaction. QCD successfully explains many experimental data, in the high-energy region, the coupling constant is small, one can use perturbative method to handle the interaction. However, in the low energy region, the coupling constant is large and the perturbative method is not applicable. So far, no other analytical methods have been developed to solve this problem. In other words, the understanding of strong interaction is still incomplete. On the other hand, the properties of mesons are dominated by the non-perturbative effects of QCD, so mesons have become an ideal laboratory for the study of strong interactions in the strongly coupled non-perturbative regime \cite{Godfrey:1998pd,Li:2004gu}. The investigation of the meson spectrum is of great scientific significance for a better understanding of the non-perturbative effects of QCD.

In this work, we will focus on the assignment of $1^{3}D_{1}$ meson nonet. In the past few years, there have been a lot of analysis on the assignment of $1^{3}D_{1}$ meson nonet, but until now, there are still many confusing aspects about this problem \cite{Liang:2013jmx}. In the recent edition of Particle Data Group, the $1^{3}D_{1}$ meson nonet assignments have been presented in the quark model \cite{ParticleDataGroup:2020ssz}. Here we list the masses and decays of $1^{3}D_{1}$ meson nonet in Tab. \uppercase\expandafter{\romannumeral1}.

%%%%%%%%%%%%%%%%%%%%%%%%%%%%%%%%%%%%%%%%%%%%%%%%%%%%%%%%%%%%%%%%%%%
\begin{table}[h]
\label{Tab:t1}
 \caption{Masses and decay widths of $1^{3}D_{1}$ meson nonet in PDG.
  $^{\dag}$The physical vector mesons is mixtures of $1^{3}D_{1}$ and $2^{3}S_{1}$.
  $^{\ddag}$This state has also been proposed as a tetraquark state.
  $\phi(???)^{\ast}$ denotes the $s\bar{s}$ member of $1^{3}D_{1}$ meson nonet and has not been observed in experiment. }
\begin{ruledtabular}
\begin{tabular}{llll}
  state                            &Mass(MeV)            &   Width(MeV)
\\
\hline
 $\mathbf{\rho(1700)}$          &$1720\pm20$         &   $250\pm100$
\\
 $\mathbf{\omega(1650)}$          &$1670\pm30$         &  $315\pm35$
\\
 $K^{*}(1680)^{\dag}$          &$1718\pm18$         &   $322\pm110$
\\
$\phi(2170)^{\ddag}$ or $\phi(???)^{\ast}$          &$2159\pm17$         &   $137\pm16$
\\
\end{tabular}
\end{ruledtabular}
\end{table}
%%%%%%%%%%%%%%%%%%%%%%%%%%%%%%%%%%%%%%%%%%%%%%%%%%%%%%%%%%%%%%%%%%%

There is some controversy about $K^{*}(1680)$ as the kaon in the $1^{3}D_{1}$ meson nonet. To specify this question, we need to briefly review the kaon state $K^{*}(1410)$ in the $2^{3}S_{1}$ meson nonet. In PDG, the $K^{*}(1410)$ as candidate for $2^{3}S_{1}$ meson nonet, but there are still some problems about this assignment. On the one hand, the mass of $K^{*}(1410)$($1414\pm15MeV$) is smaller compared with other states $\rho(1450)$ and $\omega(1420)$ of $2^{3}S_{1}$ meson nonet. In the past few years, people have analyzed the mass of $K^{*}(1410)$  with the different models, e.g. Godfrey-Isgur quark model ($\sim1580MeV$) \cite{Godfrey:1985xj}, constituent quark model ($\sim1620MeV$) \cite{Vijande:2004he}, semirelativistic potential model ($\sim1600MeV$) \cite{Brau:2002zpy} and Regge trajectories ($\sim1608MeV$) \cite{Tornqvist:1990fv}. The mass range does not support such assignment. On the other hand, the reported $\pi K$ branching fraction deviate significantly from the theoretical prediction \cite{Barnes:2002mu}.
Burakovsky also indicated the mass of $K^{\ast}(1410)$ seems too light to be the $2^3S_1$ meson nonet \cite{Burakovsky:1997ch}. In Ref. \cite{Tornqvist:1990fv}, the existence of state $K^{\ast}(1410)$ is doubted by $T\ddot{o}rnquist$. In our previous work \cite{Feng:2007zze}, we also suggested that the assignment of $K^{\ast}(1410)$ should be tested in the future experiments. Many references above suggested that the assignment of state $K^{\ast}(1410)$ should be revisited in the future. The earlier edition of PDG recommended that the $K^{\ast}(1410)$ could be replaced by the $K^{\ast}(1680)$ as the $2^3S_1$ state \cite{ParticleDataGroup:2006fqo}. If the replacement is testified reasonably in experiment, we would encounter another interesting puzzle, which state should be the candidate for kaon in the $1^3D_1$ meson nonet. The state $K^\ast(1680)$ with the mass $1717\pm27 MeV$ and full width $322 \pm 110 MeV$ is observed in the $\pi K$, $\rho K$ and $K ^{\ast}(892)\pi$ decay modes at present \cite{ParticleDataGroup:2020ssz}. However, Barnes indicated this state should have large branch fraction in $\pi K_1(1273)$ decay mode in the $^3P_0$ model \cite{Barnes:2002mu}. Similarly, $K^{*}(1680)$ mass seems too light if we accept the $\rho(1700)$ and $\omega(1650)$ as $1^{3}D_{1}$ $n\bar{n}$ states. Moreover, the $K^{\ast}(1410)$ and $K^\ast(1680)$ could be mixtures of $2^{3}S_{1}$ and $1^{3}D_{1}$ states, this situation makes the problem more interesting \cite{Pang:2017dlw}.

Apart from the kaon, the $s\bar{s}$ member of $1^{3}D_{1}$ meson nonet is even more disturbing, with the $\rho(1700)$ and $\omega(1650)$ as the well established $1^{3}D_{1}$ $n\bar{n}$ states, the $s\bar{s}$ member mass is investigated in different models, e.g. Godfrey-Isgur quark model \cite{Godfrey:1985xj}, the flux-tube model \cite{Barnes:1995hc}. Surprisingly, the $\phi(2170)$ is assigned as the the $s\bar{s}$ member of $1^{3}D_{1}$ meson nonet in PDG \cite{ParticleDataGroup:2020ssz}. Because the mass of $\phi(2170)$ deviates greatly from the conventional predictions, we have doubts about such assignment. It is also pointed out that $\phi(2170)$ proposed as a tetraquark \cite{Agaev:2019coa}. A more detailed discussion of this state can be found in Refs. \cite{Agaev:2019coa,BESIII:2020gnc}.

In this work, we will investigate the mass spectrum of $1^{3}D_{1}$ meson nonet in the framework of Regge phenomenology. In addition, the decays of $1^{3}D_{1}$ meson nonet are given. The article is organized as follows: In Sec. \uppercase\expandafter{\romannumeral2}, a brief review Regge phenomenology and the mass relations of $1^{3}D_{1}$ state are obtained. In Sec. \uppercase\expandafter{\romannumeral3}, decays of $\phi(???)$ are presented in the $^{3}P_{0}$ model, and a summary is given in Sec. \uppercase\expandafter{\romannumeral4}.

\section{Regge phenomenology and mass relations}

Regge theory originated from the analysis of scattering amplitudes in complex angular momentum space in 1959 \cite{Regge:1959mz}, and it is applied to the study of high energy particles physics. Regge theory involves almost all aspects of strong interactions, including hadron spectrum, the forces of particles, and the high energy behavior of scattering amplitudes \cite{Collins:1971ff,Chen:2016spr,Guo:2008he,Anisovich:2000kxa,Masjuan:2012gc,Pang:2019ovr,Chen:2018nnr,Chen:2021kfw,Chen:2022flh,Li:2007px}.In Regge theory, the pole of split wave amplitude in the plane of the complex angular momentum is called Regge pole, and the curve depended by it moving in the plane when energy changes is called Regge trajectory. The plots of Regge trajectories of hadrons in the $(J, M^{2})$ plane are usually called Chew-Frautschi plots (where $J$ and $M$ are respectively the total spins and the masses of the hadrons). In the past two decades, the quasilinear Regge trajectory was used for studying hadron spectra and result in reasonable description for the hadron spectroscopy \cite{Li:2004gu,Guo:2008he,Anisovich:2000kxa,Li:2007px}.

In the present work, we investigate the $1^{3}D_{1}$ meson nonet in the framework of quasilinear Regge trajectories. The quasi-linear Regge trajectory for
a meson state is usually parameterized as \cite{Li:2004gu,Guo:2008he,Li:2007px}
\begin{equation}
J=\alpha_{n\bar{n}({N^{2S+1}L_{J}})}(0)+\alpha'_{n\bar{n}({N^{2S+1}L_{J}})}M^2_{n\bar{n}({N^{2S+1}L_{J}})}
\end{equation}
\begin{equation}
J=\alpha_{n\bar{s}({N^{2S+1}L_{J}})}(0)+\alpha'_{n\bar{s}({N^{2S+1}L_{J}})}M^2_{n\bar{s}({N^{2S+1}L_{J}})}
\end{equation}
\begin{equation}
J=\alpha_{s\bar{s}({N^{2S+1}L_{J}})}(0)+\alpha'_{s\bar{s}({N^{2S+1}L_{J}})}M^2_{s\bar{s}({N^{2S+1}L_{J}})}
\end{equation}
\begin{equation}
J=\alpha_{c\bar{c}({N^{2S+1}L_{J}})}(0)+\alpha'_{c\bar{c}({N^{2S+1}L_{J}})}M^2_{c\bar{c}({N^{2S+1}L_{J}})}
\end{equation}
\begin{equation}
J=\alpha_{c\bar{s}({N^{2S+1}L_{J}})}(0)+\alpha'_{c\bar{s}({N^{2S+1}L_{J}})}M^2_{c\bar{s}({N^{2S+1}L_{J}})}
\end{equation}
\begin{equation}
J=\alpha_{c\bar{n}({N^{2S+1}L_{J}})}(0)+\alpha'_{c\bar{n}({N^{2S+1}L_{J}})}M^2_{c\bar{n}({N^{2S+1}L_{J}})}
\end{equation}
where $n$ ($n=u$ or $d$ quark), $s$ and $c$ refer to the quark constituents, $J$ and $M$ are the spin and mass of the meson state, respectively. $N$ is the radial quantum number, $L$ is orbital angular momentum. The $\alpha'$ and $\alpha$ are the slope and intercept of the Regge trajectory. For a given meson state, the intercept and slope can be expressed by the following relations.
\begin{equation}
\alpha_{n\bar{n}({N^{2S+1}L_{J}})}(0)+\alpha_{s\bar{s}({N^{2S+1}L_{J}})}(0)=2\alpha_{n\bar{s}({N^{2S+1}L_{J}})}(0)
\end{equation}
\begin{equation}
\alpha_{c\bar{c}({N^{2S+1}L_{J}})}(0)+\alpha_{n\bar{n}({N^{2S+1}L_{J}})}(0)=2\alpha_{c\bar{n}({N^{2S+1}L_{J}})}(0)
\end{equation}
\begin{equation}
\alpha_{c\bar{c}({N^{2S+1}L_{J}})}(0)+\alpha_{s\bar{s}({N^{2S+1}L_{J}})}(0)=2\alpha_{c\bar{s}({N^{2S+1}L_{J}})}(0)
\end{equation}
\begin{equation}
\frac{1}{\alpha'_{n\bar{n}({N^{2S+1}L_{J}})}}+\frac{1}{\alpha'_{s\bar{s}({N^{2S+1}L_{J}})}}=\frac{2}{\alpha'_{n\bar{s}({N^{2S+1}L_{J}})}}
\end{equation}
\begin{equation}
\frac{1}{\alpha'_{c\bar{c}({N^{2S+1}L_{J}})}}+\frac{1}{\alpha'_{n\bar{n}({N^{2S+1}L_{J}})}}=\frac{2}{\alpha'_{c\bar{n}({N^{2S+1}L_{J}})}}
\end{equation}
\begin{equation}
\frac{1}{\alpha'_{c\bar{c}({N^{2S+1}L_{J}})}}+\frac{1}{\alpha'_{s\bar{s}({N^{2S+1}L_{J}})}}=\frac{2}{\alpha'_{c\bar{s}({N^{2S+1}L_{J}})}}
\end{equation}

The intercept correlations (7), (8) and (9) were derived from in the dual-resonance model \cite{Berezinsky:1969erk}, and are satisfied in two-dimensional QCD \cite{Brower:1977as}, the dual-analytic model \cite{Kobylinsky:1978db}, and the quark bremsstrahlung model \cite{Dixit:1979mz}. The slope relations (10), (11) and (12) are obtained in the framework of topological expansion and the $q\bar{q}$-string picture of hadrons \cite{Kaidalov:1980bq}.

Combining the relations (1)-(6) and (7)-(9) , one obtains
\begin{equation}
M^{2}_{n\bar{n}({N^{2S+1}L_{J}})}\alpha'_{n\bar{n}({N^{2S+1}L_{J}})}+M^{2}_{s\bar{s}({N^{2S+1}L_{J}})}\alpha'_{s\bar{s}({N^{2S+1}L_{J}})}=2M^{2}_{n\bar{s}({N^{2S+1}L_{J}})}\alpha'_{n\bar{s}({N^{2S+1}L_{J}})}
\end{equation}
\begin{equation}
M^{2}_{n\bar{n}({N^{2S+1}L_{J}})}\alpha'_{n\bar{n}({N^{2S+1}L_{J}})}+M^{2}_{c\bar{c}({N^{2S+1}L_{J}})}\alpha'_{c\bar{c}({N^{2S+1}L_{J}})}=2M^{2}_{c\bar{n}({N^{2S+1}L_{J}})}\alpha'_{c\bar{n}({N^{2S+1}L_{J}})}
\end{equation}
\begin{equation}
M^{2}_{s\bar{s}({N^{2S+1}L_{J}})}\alpha'_{s\bar{s}({N^{2S+1}L_{J}})}+M^{2}_{c\bar{c}({N^{2S+1}L_{J}})}\alpha'_{c\bar{c}({N^{2S+1}L_{J}})}=2M^{2}_{c\bar{s}({N^{2S+1}L_{J}})}\alpha'_{c\bar{s}({N^{2S+1}L_{J}})}
\end{equation}

Apply the relations (13), (14) and (15) to the $1^{3}S_{1}$, $1^{3}P_{2}$ and $1^{3}D_{1}$ meson state, the following relations are obtained by eliminating the slopes.
Our analysis is based on these assumptions that the slopes of parity partners trajectories coincide and the slopes of ground and radial excited states are the same, which is widely used in Refs. \cite{Anisovich:2000kxa,Li:2007px,Anisovich:2001ig,Anisovich:2002us}, that is to say,
$\alpha'_{n\bar{n}({N^{3}S_{1}})}=\alpha'_{n\bar{n}({N^{3}P_{2}})} =\alpha'_{n\bar{n}({N^{3}D_{1}})}$,
 $\alpha'_{n\bar{s}({N^{3}S_{1}})}=\alpha'_{n\bar{s}({N^{3}P_{2}})} =\alpha'_{n\bar{s}({N^{3}D_{1}})}$,
 $\alpha'_{s\bar{s}({N^{3}S_{1}})}=\alpha'_{s\bar{s}({N^{3}P_{2}})} =\alpha'_{s\bar{s}({N^{3}D_{1}})}$,
 $\alpha'_{c\bar{c}({N^{3}S_{1}})}=\alpha'_{c\bar{c}({N^{3}P_{2}})} $,
 $\alpha'_{c\bar{n}({N^{3}S_{1}})}=\alpha'_{c\bar{n}({N^{3}P_{2}})} $,
 $\alpha'_{c\bar{s}({N^{3}S_{1}})}=\alpha'_{c\bar{s}({N^{3}P_{2}})} $.
\begin{equation}
\frac  { 4M^{2}_{n\bar{s}({1^{3}S_{1}})}  M^{2}_{n\bar{n}({1^{3}P_{2}})}-4M^{2}_{n\bar{n}({1^{3}S_{1}})}   M^{2}_{n\bar{s}({1^{3}P_{2}})}  }
       { M^{2}_{n\bar{n}({1^{3}S_{1}})}   M^{2}_{s\bar{s}({1^{3}P_{2}})}- M^{2}_{s\bar{s}({1^{3}S_{1}})}   M^{2}_{n\bar{n}({1^{3}P_{2}})}  }
= \frac { M^{2}_{n\bar{s}({1^{3}S_{1}})} ( M^{2}_{n\bar{n}({1^{3}P_{2}})} - M^{2}_{s\bar{s}({1^{3}P_{2}})} )  -
          M^{2}_{n\bar{s}({1^{3}P_{2}})} ( M^{2}_{n\bar{n}({1^{3}S_{1}})} - M^{2}_{s\bar{s}({1^{3}S_{1}})} )         }
        { M^{2}_{n\bar{s}({1^{3}S_{1}})}   M^{2}_{s\bar{s}({1^{3}P_{2}})}- M^{2}_{s\bar{s}({1^{3}S_{1}})}   M^{2}_{n\bar{s}({1^{3}P_{2}})}   }
\end{equation}
\begin{equation}
\frac  { 4M^{2}_{c\bar{s}({1^{3}S_{1}})}  M^{2}_{c\bar{n}({1^{3}P_{2}})}-4M^{2}_{c\bar{n}({1^{3}S_{1}})}   M^{2}_{c\bar{s}({1^{3}P_{2}})}  }
       { M^{2}_{c\bar{c}({1^{3}S_{1}})}   M^{2}_{s\bar{s}({1^{3}P_{2}})}- M^{2}_{s\bar{s}({1^{3}S_{1}})}   M^{2}_{c\bar{c}({1^{3}P_{2}})}  }
= \frac { M^{2}_{c\bar{s}({1^{3}S_{1}})} ( M^{2}_{c\bar{n}({1^{3}P_{2}})} - M^{2}_{s\bar{s}({1^{3}P_{2}})} )  -
          M^{2}_{c\bar{s}({1^{3}P_{2}})} ( M^{2}_{c\bar{n}({1^{3}S_{1}})} - M^{2}_{s\bar{s}({1^{3}S_{1}})} )         }
        { M^{2}_{c\bar{s}({1^{3}S_{1}})}   M^{2}_{s\bar{s}({1^{3}P_{2}})}- M^{2}_{s\bar{s}({1^{3}S_{1}})}   M^{2}_{c\bar{s}({1^{3}P_{2}})}   }
\end{equation}
\begin{equation}
\frac  { 4M^{2}_{n\bar{s}({1^{3}D_{1}})}  M^{2}_{n\bar{n}({1^{3}P_{2}})}-4M^{2}_{n\bar{n}({1^{3}D_{1}})}   M^{2}_{n\bar{s}({1^{3}P_{2}})}  }
       { M^{2}_{n\bar{n}({1^{3}D_{1}})}   M^{2}_{s\bar{s}({1^{3}P_{2}})}- M^{2}_{s\bar{s}({1^{3}D_{1}})}   M^{2}_{n\bar{n}({1^{3}P_{2}})}  }
= \frac { M^{2}_{n\bar{s}({1^{3}D_{1}})} ( M^{2}_{n\bar{n}({1^{3}P_{2}})} - M^{2}_{s\bar{s}({1^{3}P_{2}})} )  -
          M^{2}_{n\bar{s}({1^{3}P_{2}})} ( M^{2}_{n\bar{n}({1^{3}D_{1}})} - M^{2}_{s\bar{s}({1^{3}D_{1}})} )         }
        { M^{2}_{n\bar{s}({1^{3}D_{1}})}   M^{2}_{s\bar{s}({1^{3}P_{2}})}- M^{2}_{s\bar{s}({1^{3}D_{1}})}   M^{2}_{n\bar{s}({1^{3}P_{2}})}   }
\end{equation}
\begin{equation}
\frac  { 4M^{2}_{n\bar{s}({1^{3}D_{1}})}  M^{2}_{n\bar{n}({1^{3}S_{1}})}-4M^{2}_{n\bar{n}({1^{3}D_{1}})}   M^{2}_{n\bar{s}({1^{3}S_{1}})}  }
       { M^{2}_{n\bar{n}({1^{3}D_{1}})}   M^{2}_{s\bar{s}({1^{3}S_{1}})}- M^{2}_{s\bar{s}({1^{3}D_{1}})}   M^{2}_{n\bar{n}({1^{3}S_{1}})}  }
= \frac { M^{2}_{n\bar{s}({1^{3}D_{1}})} ( M^{2}_{n\bar{n}({1^{3}S_{1}})} - M^{2}_{s\bar{s}({1^{3}S_{1}})} )  -
          M^{2}_{n\bar{s}({1^{3}S_{1}})} ( M^{2}_{n\bar{n}({1^{3}D_{1}})} - M^{2}_{s\bar{s}({1^{3}D_{1}})} )         }
        { M^{2}_{n\bar{s}({1^{3}D_{1}})}   M^{2}_{s\bar{s}({1^{3}S_{1}})}- M^{2}_{s\bar{s}({1^{3}D_{1}})}   M^{2}_{n\bar{s}({1^{3}S_{1}})}   }
\end{equation}
\begin{equation}
\frac  { 4M^{2}_{n\bar{s}({2^{3}s_{1}})}  M^{2}_{n\bar{n}({1^{3}P_{2}})}-4M^{2}_{n\bar{n}({2^{3}S_{1}})}   M^{2}_{n\bar{s}({1^{3}P_{2}})}  }
       { M^{2}_{n\bar{n}({2^{3}S_{1}})}   M^{2}_{s\bar{s}({1^{3}P_{2}})}- M^{2}_{s\bar{s}({2^{3}S_{1}})}   M^{2}_{n\bar{n}({1^{3}P_{2}})}  }
= \frac { M^{2}_{n\bar{s}({2^{3}S_{1}})} ( M^{2}_{n\bar{n}({1^{3}P_{2}})} - M^{2}_{s\bar{s}({1^{3}P_{2}})} )  -
          M^{2}_{n\bar{s}({1^{3}P_{2}})} ( M^{2}_{n\bar{n}({2^{3}S_{1}})} - M^{2}_{s\bar{s}({2^{3}S_{1}})} )         }
        { M^{2}_{n\bar{s}({2^{3}S_{1}})}   M^{2}_{s\bar{s}({1^{3}P_{2}})}- M^{2}_{s\bar{s}({2^{3}S_{1}})}   M^{2}_{n\bar{s}({1^{3}P_{2}})}   }
\end{equation}
\begin{equation}
\frac  { 4M^{2}_{n\bar{s}({2^{3}s_{1}})}  M^{2}_{n\bar{n}({1^{3}S_{1}})}-4M^{2}_{n\bar{n}({2^{3}S_{1}})}   M^{2}_{n\bar{s}({1^{3}S_{1}})}  }
       { M^{2}_{n\bar{n}({2^{3}S_{1}})}   M^{2}_{s\bar{s}({1^{3}S_{1}})}- M^{2}_{s\bar{s}({2^{3}S_{1}})}   M^{2}_{n\bar{n}({1^{3}S_{1}})}  }
= \frac { M^{2}_{n\bar{s}({2^{3}S_{1}})} ( M^{2}_{n\bar{n}({1^{3}S_{1}})} - M^{2}_{s\bar{s}({1^{3}S_{1}})} )  -
          M^{2}_{n\bar{s}({1^{3}S_{1}})} ( M^{2}_{n\bar{n}({2^{3}S_{1}})} - M^{2}_{s\bar{s}({2^{3}S_{1}})} )         }
        { M^{2}_{n\bar{s}({2^{3}S_{1}})}   M^{2}_{s\bar{s}({1^{3}S_{1}})}- M^{2}_{s\bar{s}({2^{3}S_{1}})}   M^{2}_{n\bar{s}({1^{3}S_{1}})}   }
\end{equation}

In the PDG, the $1^{3}S_{1}$ meson multiplet $\rho(770)$, $K^{*}(892)$, $J/\psi(1S)$, $D^{*}$ and $D_{s}^{*\pm}$, $1^{3}P_{2}$ meson multiplet $a_{2}(1320)$, $K_{2}^{*}(1430)$, $\chi_{c2}(1P)$, $D_{2}^{*}(2420)$  and $1^{3}D_{1}$ meson state $\rho(1700)$ are well established. Inserting the masses of these states to the relations (16)-(21), we obtain the following meson masses $M_{n\bar{s}({1^{3}D_{1}})}=1835.1\pm6.3 MeV$, $M_{s\bar{s}({1^{3}D_{1}})}=1944.1\pm6.6 MeV$, $M_{n\bar{s}({2^{3}S_{1}})}=1562.9 \pm9.1MeV$, $M_{s\bar{s}({2^{3}S_{1}})}=1655.9\pm9.7 MeV$.

Next, we discuss the masses of radial excited state of $1^{3}D_{1}$ meson nonet. In the present work, considering the fact that the ground and the radial excitations have the same slopes, we can have the following relations from (1)-(3).
\begin{equation}
M^{2}_{n\bar{n}({N^{2S+1}L_{J}})}\alpha'_{n\bar{n}}-M^{2}_{n\bar{n}({N'^{2S+1}L_{J}})}\alpha'_{n\bar{n}}=\alpha_{n\bar{n}({N'^{2S+1}L_{J}})}(0)-\alpha_{n\bar{n}({N^{2S+1}L_{J}})}(0)
\end{equation}

\begin{equation}
M^{2}_{s\bar{s}({N^{2S+1}L_{J}})}\alpha'_{s\bar{s}}-M^{2}_{s\bar{s}({N'^{2S+1}L_{J}})}\alpha'_{s\bar{s}}=\alpha_{s\bar{s}({N'^{2S+1}L_{J}})}(0)-\alpha_{s\bar{s}({N^{2S+1}L_{J}})}(0)
\end{equation}

\begin{equation}
M^{2}_{n\bar{s}({N^{2S+1}L_{J}})}\alpha'_{n\bar{s}}-M^{2}_{n\bar{s}({N'^{2S+1}L_{J}})}\alpha'_{n\bar{s}}=\alpha_{n\bar{s}({N'^{2S+1}L_{J}})}(0)-\alpha_{n\bar{s}({N^{2S+1}L_{J}})}(0)
\end{equation}

$N$ and $N'$ are the radial quantum numbers, $N'=1$ is ground state. Based on the assumption that the dispersion of $\alpha_{q\bar{q'}N}(0)-\alpha_{q\bar{q'}1}(0)$ is flavor independent, the expression $\alpha_{n\bar{n}({N'^{2S+1}L_{J}})}(0)-\alpha_{n\bar{n}({N^{2S+1}L_{J}})}(0)$ can be simplified to $N'-N$, which is used in Refs.  \cite{Anisovich:2000kxa,Anisovich:2002us,Anisovich:2003tm}. However, by introducing the latest existing experimental data, one can find this assumption will lead to large deviation. After a comprehensive phenomenological analysis of some well-established mesons, Filipponi et al. pointed out that the values of
 $\alpha_{n\bar{n}({N'^{2S+1}L_{J}})}(0)-\alpha_{n\bar{n}({N^{2S+1}L_{J}})}(0)$, $\alpha_{s\bar{s}({N'^{2S+1}L_{J}})}(0)-\alpha_{s\bar{s}({N^{2S+1}L_{J}})}(0)$ and $\alpha_{n\bar{s}({N'^{2S+1}L_{J}})}(0)-\alpha_{n\bar{s}({N^{2S+1}L_{J}})}(0)$ depend on the constituent quark masses through the combination $m_{i} + m_{j}$ ($m_{i}$ and $m_{j}$ are the constituent masses of quark and antiquark) in Refs. \cite{Filipponi:1997hb,Filipponi:1997vf}. In this work, we introduce parameters $f_{i\overline{j}}(m_{i}+m_{j})$ into relations (22)-(24) and result in the following relations (25)-(27)\cite{Li:2007px,Liu:2010zzd}.
\begin{equation}
M^{2}_{n\bar{n}({N^{2S+1}L_{J}})}-M^{2}_{n\bar{n}({N'^{2S+1}L_{J}})}=\frac{(N-N')}{\alpha'_{n\bar{n}}}(1+f_{n\bar{n}}(m_{n}+m_{n}))
\end{equation}

\begin{equation}
M^{2}_{s\bar{s}({N^{2S+1}L_{J}})}-M^{2}_{s\bar{s}({N'^{2S+1}L_{J}})}=\frac{(N-N')}{\alpha'_{s\bar{s}}}(1+f_{s\bar{s}}(m_{s}+m_{s}))
\end{equation}

\begin{equation}
M^{2}_{n\bar{s}({N^{2S+1}L_{J}})}-M^{2}_{n\bar{s}({N'^{2S+1}L_{J}})}=\frac{(N-N')}{\alpha'_{n\bar{s}}}(1+f_{n\bar{s}}(m_{n}+m_{s}))
\end{equation}

In relations (25)-(27), we take $m_{n}=0.29GeV$, $m_{s}=0.45GeV$, $\alpha'_{n\bar{n}}=0.8830GeV^{-2}$, $\alpha'_{n\bar{s}}=0.8493GeV^{-2}$, $\alpha'_{s\bar{s}}=0.8181GeV^{-2}$ as input \cite{Li:2007px,Liu:2010zzd}. With the aid of the masses $\rho(770)$, $\rho(1450)$, $M_{n\bar{s}}(1^{3}S_{1})$, $M_{n\bar{s}}(2^{3}S_{1})$, $M_{s\bar{s}}(2^{3}S_{1})$, the parameters $f_{n\bar{n}}$, $f_{n\bar{n}}$, $f_{n\bar{n}}$ are determined to be
\begin{equation}
f_{n\bar{n}}=0.63\pm0.12GeV^{-1}, f_{n\bar{s}}=0.54\pm0.04GeV^{-1}, f_{s\bar{s}}=0.46\pm0.03GeV^{-1}
\end{equation}

Based on the relations (25)-(27), the radial excitation masses of the $N^{3}D_{1}$ multiplet can be estimated. Our predictions and those given by other references are listed in Tab. \uppercase\expandafter{\romannumeral2} and Fig. 1.

%%%%%%%%%%%%%%%%%%%%%%%%%%%%%%%%%%%%%%%%%%%%%%%%%%%%%%%%%%%%%%%%%%%
\begin{table}[h]
\label{table} \caption{The radial excitation masses of the $N^{3}D_{1}$ multiplet. (in units of MeV)  The masses used as input for our calculation
are shown in boldface. }
\begin{ruledtabular}
\begin{tabular}{lllllllllll}
  state                     &           &$M_{n\bar{n}}$ &          &           &$M_{n\bar{s}}$&          &          &$M_{s\bar{s}}$ &        &
\\
\hline
 $Reference$                &   $N=1$            &   $N=2$      &   $N=3$   &   $N=1$   &   $N=2$   & $N=3$     &   $N=1$     &   $N=2$       & $N=3$ &
\\
 $Present work$             &   $\mathbf{{1720}\pm20}$  &   $2112.2\pm23.7$   & $2459.4\pm32.3$  & $1823.6\pm6.2$  &   $2239.4\pm8.2$  &$2581.3\pm11.9$    &   $1944.1\pm6.6$  & $2346.4\pm8.8$    &  $2689.2\pm12.9$&
\\
 Ref. \cite{Godfrey:1985xj}  &    1660            &  2150        &            &  1780    &  2250     &           &   1880      &       &        &
\\
 Ref. \cite{Xiao:2019qhl}  &           &              &            &           &                   &           &  $1883$     &   $2342$      & $2732$  &
\\
 Ref. \cite{Li:2020xzs}    &           &              &            &           &                   &           &  $1809$     &   $2272$      & $2681$  &
\\
 Ref. \cite{Pang:2019ttv} &           &              &            &           &                   &           &  $1869$     &   $2276$      & $2593$  &
\\
 Ref. \cite{Ebert:2009ub} &   1557    &  1895        & 2168       &           &                   &           &  $1845$     &   $2258$      & $2607$  &
\\
 Ref. \cite{Pang:2018gcn} &   1646    &  2048       & 2364       &   1765        &                   &           &  $1845$     &   $2258$      & $2607$  &
\\
\end{tabular}
\end{ruledtabular}
\end{table}
%%%%%%%%%%%%%%%%%%%%%%%%%%%%%%%%%%%%%%%%%%%%%%%%%%%%%%%%%%%%%%%%%%%

%%%%%%%%%%%%%%%%%%%%%%%%%%%%%%%%%%%%%%%%%%%%%%%%%%%%%%%%%%%%%%%%
%%%  Figure
\begin{figure}[htbp]
\begin{center}
\includegraphics[width=0.5\textwidth]{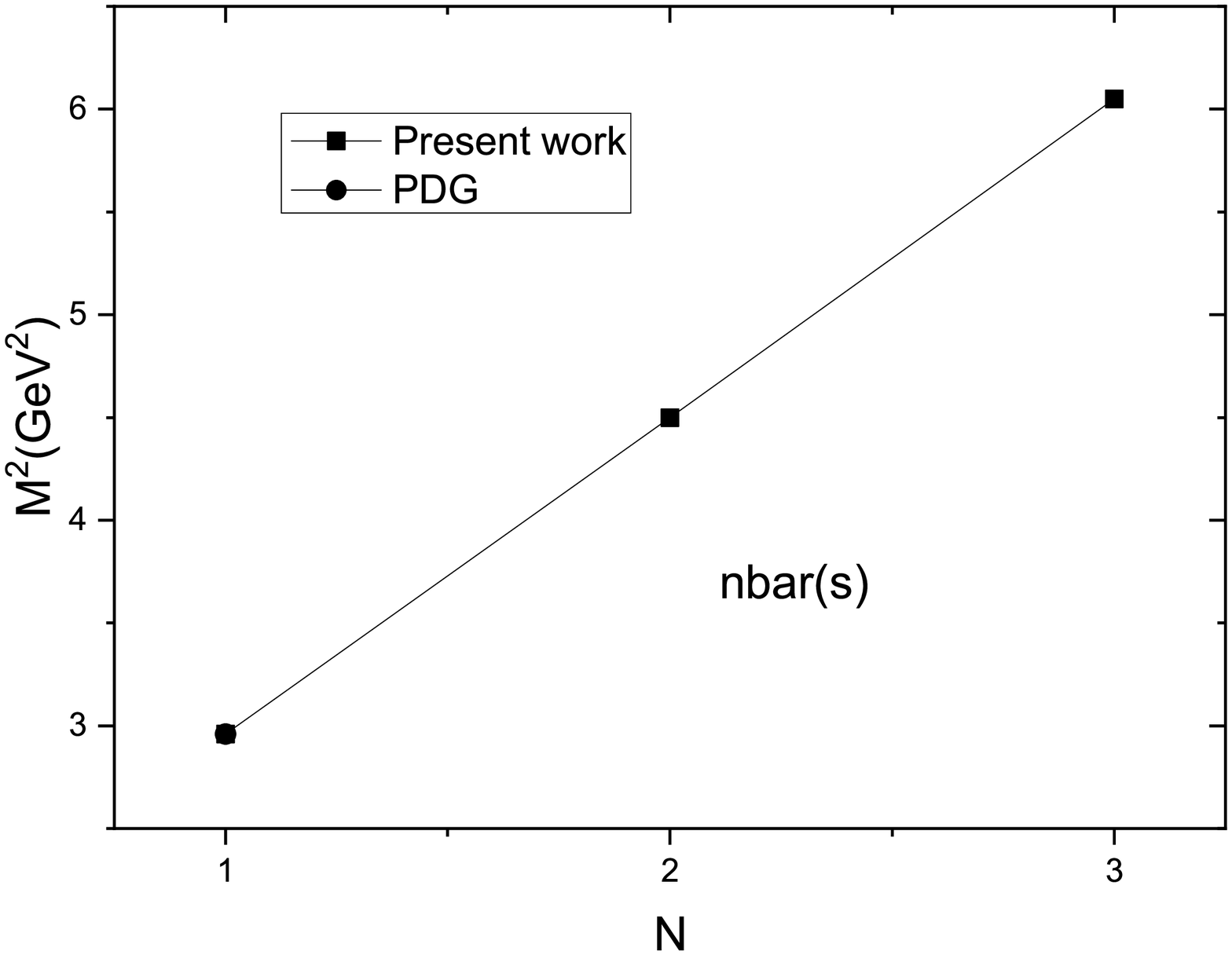}
\includegraphics[width=0.5\textwidth]{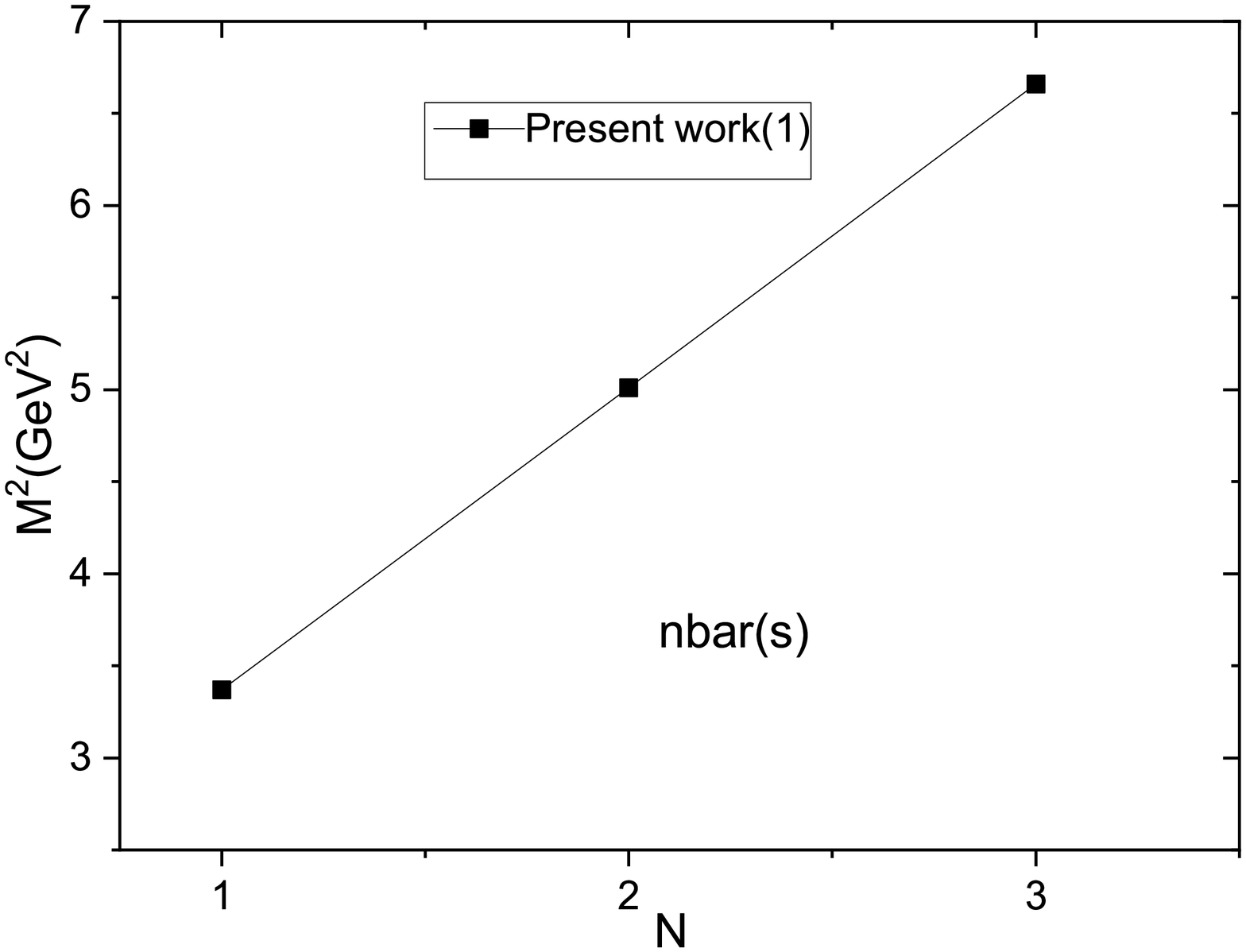}
\includegraphics[width=0.5\textwidth]{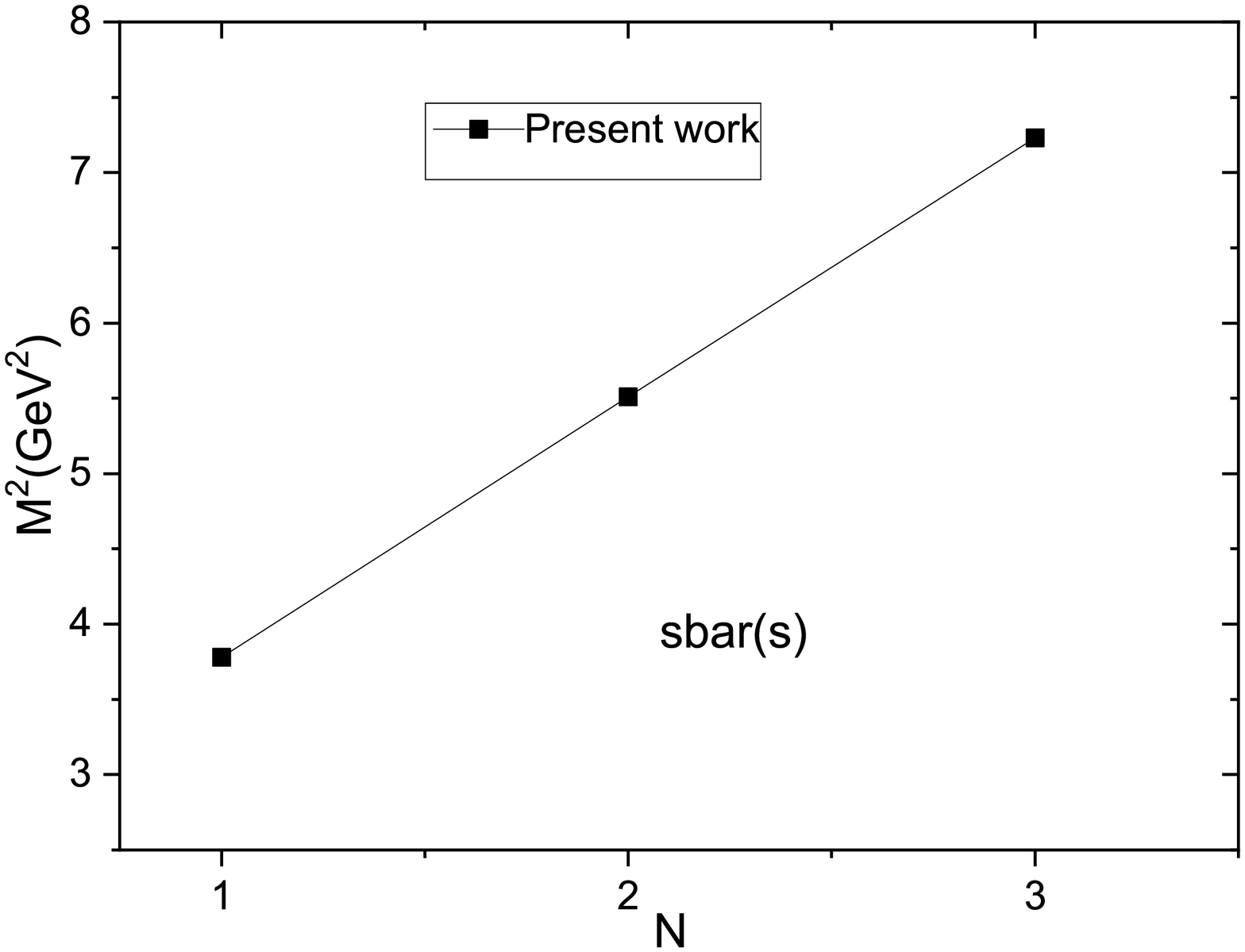}
\caption{The radial excitation of $N^{3}D_{1}$ meson nonet.}
\end{center}
\end{figure}
%%%%%%%%%%%%%%%%%%%%%%%%%%%%%%%%%%%%%%%%%%%%%%%%%%%%%%%%%%%%%%%%

\section{Decays of $1^{3}D_{1}$ meson nonet }
In addition to the analysis of mass spectrum, the decays of $1^{3}D_{1}$ meson nonet are investigated in the $^{3}P_{0}$ model. The $^{3}P_{0}$ model was first proposed by Micu \cite{Micu:1968mk}, and was applied to OZI-allowed strong decays of mesons. In the 1970s, Le Yaouanc et al. further developed the $^{3}P_{0}$ model \cite{LeYaouanc:1972vsx,LeYaouanc:1973ldf}. In this model, the decay process $A\rightarrow BC$ occurs when the quark-antiquark pair produces a state suitable for quark rearrangement from the vacuum. To this day, the model is widely used to calculate the decay amplitude and decay branch ratios of hadrons and achieved very good results \cite{Gui:2018rvv,Li:2021qgz,Hao:2019fjg,Pan:2016bac,Li:2008mza,Li:2009rka,Lu:2014zua}.

In Refs. \cite{Ackleh:1996yt,Barnes:1996ff}, strong decay amplitudes and partial widths are provided in detail. For the process $A\rightarrow BC$, the partial width is expressed as
\begin{equation}
\Gamma_{A \rightarrow B C}=2 \pi \frac{P E_{B} E_{C}}{M_{A}} \sum_{L S}\left( M_{LS}\right)^{2}
\end{equation}
with
$$
M_{LS}=\frac{\gamma}{\pi^{1 / 4} \beta^{1 / 2}} \xi_{L S}(\frac{P}{\beta}) e^{-P^{2} / 12 \beta^{2}}
$$

$$
P=\frac{\left[\left(M_{A}^{2}-\left(M_{B}+M_{C}\right)^{2}\right)\left(M_{A}^{2}-\left(M_{B}-M_{C}\right)^{2}\right)\right]^{1 / 2}}{2 M_{A}}
$$

$$
E_{B}=\frac{M_{A}^{2}-M_{C}^{2}+M_{B}^{2}}{2 M_{A}}
$$

$$
E_{B}=\frac{M_{A}^{2}+M_{C}^{2}-M_{B}^{2}}{2 M_{A}}
$$

where $P$ is the decay momentum, $E_{B}$ and $E_{C}$ are the energies of meson $B$ and $C$, $M_{A}$, $M_{B}$ are the masses meson $A$ and $B$. The decay amplitude$ M_{LS}$ is proportional to polynomial $\xi_{L S}(\frac{P}{\beta})$, which is related to decay channels and can be obtained in Refs. \cite{Ackleh:1996yt,Barnes:1996ff}. In this work, we take $\beta=0.4GeV$ and $\gamma=0.4$ as input, which is used in Refs.\cite{Ackleh:1996yt,Barnes:1996ff}. From the relation (29), the decays of $\omega(1650)$, $\rho(1700)$ and $\phi(???)$ are provided in Tab. \uppercase\expandafter{\romannumeral3},  the decays of $\omega(1650)$, $\rho(1700)$, $\phi(???)$ and $1^{3}D_{1}(n\bar{s})$ states are provided in Tab. \uppercase\expandafter{\romannumeral3} and Tab. \uppercase\expandafter{\romannumeral4}. In Tab. \uppercase\expandafter{\romannumeral3}, the partial width of the $KK_{1}(1270)$ depends on the mixing of $K_{1A}$ and $K_{1B}$ states, $ K_{1}(1270)=K_{1}\left({1}^{1}P_{1}\right) \cos \theta_{K}+K_{1}\left({1}^{3} P_{1}\right) \sin \theta_{K}$, the $\theta_{K}$ denotes the mixing angle. The mixing angle is investigated in the Refs.\cite{Divotgey:2013jba, Blundell:1995au,Pang:2017dlw}. In the present work, we take $\theta_{K}=45^{o}$ as input parameters \cite{Blundell:1995au,Pang:2017dlw}. Moreover, the decay amplitude$ M_{LS}$ and polynomial $\xi_{L S}(\frac{P}{\beta})$ are listed in the Appendix A.

%%%%%%%%%%%%%%%%%%%%%%%%%%%%%%%%%%%%%%%%%%%%%%%%%%%%%%%%%%%%%%%%%%%
\begin{table}[h]
\label{table} \caption{Strong decay properties for the $\omega(1650)$, $\rho(1700)$ and $1^{3}D_{1}(s\bar{s})$ ($\phi(???)$) states. (in units of MeV) }
\begin{ruledtabular}
\begin{tabular}{lllllllllll}
  Decay mode                                   &Present work & Ref. \cite{Barnes:2002mu}  & Ref. \cite{Li:2020xzs}   &  Ref. \cite{Pang:2019ttv}     & Ref. \cite{Piotrowska:2017rgt}
\\
\hline
$\rho(1700) \rightarrow \pi \pi$            &   45.6     & 48                    &   &  &
\\
$\rho(1700) \rightarrow \omega \pi$         &   35.1     & 35                    &   &  &
\\
$\rho(1700) \rightarrow \rho \eta$          &   31.9     & 16                   &   &  &
\\
$\rho(1700) \rightarrow h_{1}(1170) \pi$    &   121.1     & 124                    &   &  &
\\
$\omega(1650) \rightarrow \rho \pi$         &   86.1     & 101                    &   &  &
\\
$\omega(1650) \rightarrow \omega \eta$      &   28.7    & 13                   &   &  &
\\
$\omega(1650) \rightarrow b_{1}(1235) \pi$  &   123.8     & 371                   &   &  &
\\
 $\phi(???) \rightarrow K K$                   &   55.3      & 65                     & 30.5   & 40.8  & $104\pm28$
\\
 $\phi(???) \rightarrow K^{*}(892) K^{*}(892)$ &   91.1       & 5                       & 1.02   & 11.5  &
\\
 $\phi(???) \rightarrow K K^{*}(892)$          &  81.2      & 75                     & 42.0   & 57.8  &  $260\pm109$
\\
 $\phi(???) \rightarrow K K_{1}(1270)$         &  375.6    & 478                    & 620    & 423  &
\\
 $\phi(???) \rightarrow \eta\phi $             &   37.2     & 29                     & 13.2   & 13.6  &  $67\pm28$
\\
\end{tabular}
\end{ruledtabular}
\end{table}
%%%%%%%%%%%%%%%%%%%%%%%%%%%%%%%%%%%%%%%%%%%%%%%%%%%%%%%%%%%%%%%%%%%

%%%%%%%%%%%%%%%%%%%%%%%%%%%%%%%%%%%%%%%%%%%%%%%%%%%%%%%%%%%%%%%%%%%
\begin{table}[h]
\label{table} \caption{ Strong decay properties for the $1^{3}D_{1}(n\bar{s})$ state. (in units of MeV) }
\begin{ruledtabular}
\begin{tabular}{lllllllllll}
  Decay mode                                   &Present work & Decay mode                                 & Present work   & Decay mode                                 & Present work
\\
\hline
 $1^{3}D_{1}(n\bar{s})\rightarrow \omega K $        & 9.7    & $1^{3}D_{1}(n\bar{s})\rightarrow \pi K $        & 38.2   & $1^{3}D_{1}(n\bar{s})\rightarrow \phi K $         & 14.3
\\
 $1^{3}D_{1}(n\bar{s})\rightarrow \omega K^{*}$    & 46.2     & $1^{3}D_{1}(n\bar{s})\rightarrow \eta K $       & 1.6    & $1^{3}D_{1}(n\bar{s})\rightarrow \pi K^{*} $      & 27.3
\\
$1^{3}D_{1}(n\bar{s})\rightarrow h_{1} K $         & 4.3       & $1^{3}D_{1}(n\bar{s})\rightarrow \eta' K $      & 48.1   & $1^{3}D_{1}(n\bar{s})\rightarrow \eta K^{*}$      & 129.5
\\
$1^{3}D_{1}(n\bar{s})\rightarrow \pi K_{1}(1270)$  & 100.0     & $1^{3}D_{1}(n\bar{s})\rightarrow \rho K $       & 29.0   & $1^{3}D_{1}(n\bar{s})\rightarrow \eta K^{*}$      &  57.3
\\
\end{tabular}
\end{ruledtabular}
\end{table}
%%%%%%%%%%%%%%%%%%%%%%%%%%%%%%%%%%%%%%%%%%%%%%%%%%%%%%%%%%%%%%%%%%%
\section{Summary}
On the basis of the introduction in Sec. \uppercase\expandafter{\romannumeral1}, the assignment of the $1^{3}D_{1}$ meson nonet is not clear. In this paper, we established new mass relations which related the masses of meson multiplet in the framework of Regge phenomenology. Inserting the corresponding meson masses, we investigate the mass spectrum of the $1^3D_1$ meson nonet. In our work, the mass of  $1^{3}D_{1}(s\bar{s})$ is determined to be $1944.1MeV$, which is consistent with the prediction from Ref. \cite{Piotrowska:2017rgt}. Apart from mass range, the strong decay properties of the $1^{3}D_{1}(s\bar{s})$ are presented. There are various indications that further researches on the $1^{3}D_{1}(s\bar{s})$ is necessary in the future.  Our results may provide useful mass range for the phenomenological study.

\appendix
\section{the decay amplitude $ M_{LS}$ and polynomial $\xi_{L S}(\frac{P}{\beta})$ for the $1^{3}D_{1}$ meson nonet decay in $^{3}P_{0}$ model.}

$$
\xi_{{LS}{(^{3}D_{1} \rightarrow ^{1}S_{0}+ ^{1}S_{0})}}=-\sqrt{\frac{5}{12}}\frac{2^{13 / 2}}{3^{4}} x\left(1-\frac{2}{15} x^{2}\right)    \qquad ^{1} \mathrm{P}_{1}
$$

$$
\xi_{{LS}{(^{3}D_{1} \rightarrow ^{3}S_{1}+ ^{1}S_{0})}}=-\sqrt{\frac{5}{24}}\frac{2^{13 / 2}}{3^{4}} x\left(1-\frac{2}{15} x^{2}\right)    \qquad ^{3} \mathrm{P}_{1}
$$

$$
\xi_{{LS}{(^{3}D_{1} \rightarrow ^{3}S_{1}+ ^{3}S_{1})}}=\left\{\begin{array}{cc}
-\frac{\sqrt{5}}{6} \frac{2^{13 / 2}}{3^{4}} x\left(1-\frac{2}{15} x^{2}\right)  & \qquad{ }^{1} \mathrm{P}_{1} \\
0 & \qquad{ }^{3} \mathrm{P}_{1} \\
\frac{1}{6} \frac{2^{13 / 2}}{3^{4}} x\left(1-\frac{2}{15} x^{2}\right)  & \qquad{ }^{5} \mathrm{P}_{1} \\
\sqrt{\frac{28}{5}} \frac{2^{6}}{3^{9 / 2} 5^{1 / 2} 7^{1 / 2}} x^{3} & \qquad{ }^{5} \mathrm{~F}_{1}
\end{array}\right.
$$

$$
\xi_{{LS}{(^{3}D_{1} \rightarrow ^{1}P_{1}+ ^{1}S_{0})}}=\left\{\begin{array}{cc}
\frac{2^{6} 5^{1 / 2}}{3^{4}}\left(1-\frac{5}{18} x^{2}+\frac{1}{135} x^{4}\right) & \qquad{ }^{3} \mathrm{~S}_{1} \\
\frac{2^{15 / 2}}{3^{6} 5^{1 / 2}} x^{2}\left(1-\frac{1}{6} x^{2}\right) & \qquad{ }^{3} \mathrm{D}_{1}
\end{array}\right.
$$

$$
M^{2}_{{LS}{(^{3}D_{1} \rightarrow ^{1}S_{0}+ ^{1}S_{0})}}=\frac{\gamma}{\sqrt{\pi} \beta}\frac{5*2^{11}}{3^{9}} \left(\frac{P}{\beta}-\frac{2P^{3}}{15\beta^{3}} \right)^{2}   e^{-P^{2}/6\beta^{2}}
$$

$$
M^{2}_{{LS}{(^{3}D_{1} \rightarrow ^{3}S_{1}+ ^{1}S_{0})}}=\frac{\gamma}{\sqrt{\pi} \beta}\frac{5*2^{10}}{3^{9}} \left(\frac{P}{\beta}-\frac{2P^{3}}{15\beta^{3}} \right)^{2}   e^{-P^{2}/6\beta^{2}}
$$

$$
M^{2}_{{LS}{(^{3}D_{1} \rightarrow ^{3}S_{1}+ ^{3}S_{1})}}=\frac{\gamma}{\sqrt{\pi} \beta} \frac{5*2^{11}}{3^{10}} \left(\frac{P}{\beta}-\frac{2P^{3}}{15\beta^{3}} \right)^{2}   e^{-P^{2}/6\beta^{2}}
$$
$$+\frac{\gamma}{\sqrt{\pi} \beta}\frac{2^{11}}{3^{10}} \left(\frac{P}{\beta}-\frac{2P^{3}}{15\beta^{3}} \right)^{2}   e^{-P^{2}/6\beta^{2}}+\frac{\gamma}{\sqrt{\pi} \beta}\frac{2^{14}}{25*3^{9}} \left(\frac{P^{6}}{\beta^{6}} \right)   e^{-P^{2}/6\beta^{2}}
$$

$$
M^{2}_{{LS}{(^{3}D_{1} \rightarrow ^{1}P_{1}+ ^{1}S_{0})}}=\frac{\gamma}{\sqrt{\pi} \beta}\frac{2^{11}}{5*3^{8}} \left(\frac{P^{2}}{\beta^{2}}-\frac{P^{4}}{21\beta^{4}}\right)^{2}
e^{-P^{2}/6\beta^{2}}
$$
$$
+\frac{\gamma}{\sqrt{\pi} \beta}\frac{2^{13}}{245*3^{11}} \left(\frac{P^{8}}{\beta^{8}}\right)   e^{-P^{2}/6\beta^{2}}
$$

\end{document}